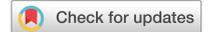

# scientific reports

**OPEN**

# Bespoke magnetic field design for a magnetically shielded cold atom interferometer

P. J. Hobson[1,2], J. Vovrosh[1], B. Stray[1], M. Packer[2], J. Winch[1], N. Holmes[2], F. Hayati[1], K. McGovern[1], R. Bowtell[2], M. J. Brookes[2], K. Bongs[1], T. M. Fromhold[1✉] & M. Holynski[1]

Quantum sensors based on cold atoms are being developed which produce measurements of unprecedented accuracy. Due to shifts in atomic energy levels, quantum sensors often have stringent requirements on their internal magnetic field environment. Typically, background magnetic fields are attenuated using high permeability magnetic shielding, with the cancelling of residual and introduction of quantisation fields implemented with coils inside the shield. The high permeability shield, however, distorts all magnetic fields, including those generated inside the sensor. Here, we demonstrate a solution by designing multiple coils overlaid on a 3D-printed former to generate three uniform and three constant linear gradient magnetic fields inside the capped cylindrical magnetic shield of a cold atom interferometer. The fields are characterised in-situ and match their desired forms to high accuracy. For example, the uniform transverse field, $B_x$, deviates by less than 0.2% over more than 40% of the length of the shield. We also map the field directly using the cold atoms and investigate the potential of the coil system to reduce bias from the quadratic Zeeman effect. This coil design technology enables targeted field compensation over large spatial volumes and has the potential to reduce systematic shifts and noise in numerous cold atom systems.

Cold atoms have been used as the basis for ultra-sensitive metrological tools in fundamental research, including testing the weak equivalence principle[1], and measuring physical quantities like the fine structure constant[2] and Electric Dipole Moment (EDM) of fundamental particles[3]. This sensitivity has prompted research to focus on moving cold atom sensors out of the laboratory to address real-world applications[4], such as inertial sensing[5], timekeeping[6], and gravity sensing[7–9], including in maritime[10] and airborne[11,12] settings. The operation of these sensors requires specific, well-controlled, magnetic field profiles to be applied while preparing the atom cloud and during measurement. Deviations from these ideal field profiles, either due to their imperfect generation or background magnetic fields, reduce measurement accuracy. In atomic preparation, for example, background magnetic fields can limit the temperature of atomic clouds in Magneto–Optical Traps (MOTs)[13]. Cold atom interferometry experiments testing the universality of free fall using atoms in magnetically sensitive sub-levels are highly sensitive to spatial deviations in the magnetic field over the trajectory through which the atoms fall[14]. If magnetic fields cause the atom fall trajectory to deviate off the axis of the sensor, they can cause the system to become susceptible to the Coriolis force[15]. Even in atom interferometry experiments using atoms in magnetically insensitive sub-levels, magnetic field deviations may induce systematic bias due to the quadratic Zeeman effect[16,17]. Although this can be mitigated by the use of a launch configuration[18] or using Raman wave vector reversal[19], it cannot be cancelled totally due to the spatial non-overlap of the interference paths during interferometry[20,21]. Other effects caused by deviations from ideal field profiles include induced bias in cold atom clocks[22,23], reduced accuracy of optically pumped magnetometers[24–26], and atomic depolarisation and Larmor frequency shifts in EDM experiments[27]. These issues are compounded when cold atom sensors are operated outside the laboratory in environments where large background fields vary both spatially and temporally.

Many cold atom sensors are enclosed with cylindrical passive magnetic shields to attenuate background magnetic fields. Passive magnetic shields are typically made of high permeability materials like mumetal[28]. However, mumetal has a crystalline structure which is damaged by external physical impacts and requires degaussing to prevent it becoming magnetised. A particular concern in portable devices is to reduce the system size, weight, and cost, while maintaining instrument sensitivity. This can be achieved by replacing some passive shielding with field coils. Compensation coils inside the magnetic shield, often based on Helmholtz pairs, are used to provide

nature portfolio

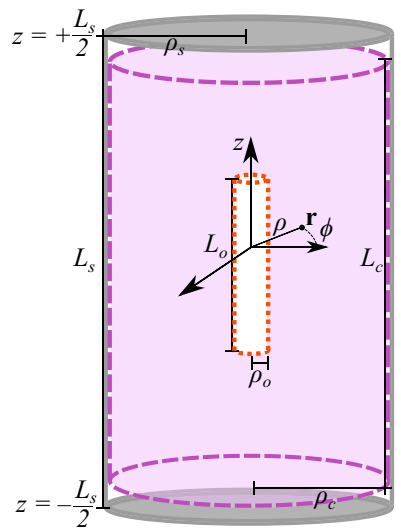

**Figure 1.** Current flow, **J**, on a current-carrying coil of radius $\rho_c$ and length $L_c$ (purple dashed), placed symmetrically inside a co-axial closed cylindrical magnetic shield of radius $\rho_s$ and length $L_s$ (grey solid), contributes to an optimal magnetic field profile in a cylindrical optimisation region of radius $\rho_o$ and length $L_o$ (red dotted). The position, **r**, is displayed in cylindrical coordinates, $(\rho, \phi, z)$.

targeted cancellation of residual magnetic fields. Additional coils generate bias fields required to define the quantisation axis for a measurement. However, combining high permeability shielding with field-generating coils is problematic because the fields generated by coils are distorted by the shield[29,30]. Some formulations accounting for electromagnetic coupling with high permeability shielding have focused on optimising discrete coil structures, i.e., the separation and extent of wire loops and saddles[31–34]. These structures are flexible but may be limited when accurate magnetic fields over large volumes. Recently, target field coil design methods[35–38] have been developed in which a complete set of orthogonal current flows is optimised on a conducting surface inside a magnetic shield, accounting directly for the electromagnetic coupling with the shield[39–43]. Rather than being confined to discrete geometries, the current is free to flow anywhere on the conducting surface. By discretising the optimal current flow pattern into a wire configuration[36–38], these methods enable optimal magnetic fields to be generated over large spatial regions.

In this work, we design, build, and characterise a coil system that generates multiple optimised magnetic fields inside a quantum sensor based on cold atoms. This sensor is a cold atom interferometer designed to measure gravitational acceleration. First, we apply a sub-set of the method in Ref.[40] to solve exactly for the magnetic field generated by a current flowing on the surface of a symmetrically placed open cylinder inside a capped cylindrical high permeability magnetic shield. We then design and manufacture multiple wire patterns on a single 3D-printed coil former using this theoretical model. The magnetic field profiles generated by these wire patterns are selected to provide a quantisation axis and compensation field during the interferometry sequence. We characterise the coil system both with and without the atomic package and explore the potential of the coil system to reduce bias due to the quadratic Zeeman effect.

## Results

**Model.** To determine the optimal coil geometry we consider a simple model of a coil inside the magnetic shield of a cold atom interferometer, shown in Fig. 1. We wish to optimise current flow on the open cylindrical coil of radius $\rho_c$ and length $L_c$ on which an arbitrary static current density flows. This current density, $\mathbf{J}(\mathbf{r}') = J_\phi(\phi', z')\,\hat{\boldsymbol{\phi}} + J_z(\phi', z')\,\hat{\mathbf{z}}$, contains azimuthal, $J_\phi(\phi', z')$, and axial, $J_z(\phi', z')$, components. The coil is placed symmetrically inside the co-axial cylindrical high permeability magnetic shield of radius $\rho_s$ and length $L_s$ that is closed by planar end-caps located at $z = \pm L_s/2$.

The magnetic field relates to the magnetic field strength, **H**, and the magnetisation of the shield, **M**, by $\mathbf{B} = \mu_0(\mathbf{H} + \mathbf{M})$, where $\mu_0$ is the magnetic permeability of free space. By casting the magnetisation of the magnetic shield as an equivalent pseudocurrent density, $\tilde{\mathbf{J}} = \nabla \times \mathbf{M}$, which can then be related to the vector potential, $\mathbf{B} = \nabla \times \mathbf{A}$, and Ampere's Law, $\mathbf{J} = \nabla \times \mathbf{H}$, we can generate the Poisson equation

$$\nabla^2 \mathbf{A}(\mathbf{r}) = -\mu_0\left(\mathbf{J}(\mathbf{r}') + \tilde{\mathbf{J}}(\mathbf{r}')\right). \tag{1}$$

Equation (1) can then be solved using a Green's function[44] expansion of the magnetic vector potential subject to boundary conditions encoding the electromagnetic coupling with the magnetic shield[40,41]. We generate these boundary conditions by considering the interface conditions[44] in the limit where the shield is a perfect magnetic conductor, i.e., the relative permeability of the shield $\mu_r \to \infty$. In this case, at the inner shield wall, the magnetic flux must flow normally to the shield's surface:

$$B_\rho \Big|_{z=\pm L_s/2} = 0\,, \qquad B_\phi \Big|_{z=\pm L_s/2,\ \rho=\rho_s} = 0\,, \qquad B_z \Big|_{\rho=\rho_s} = 0\,. \tag{2}$$

The solution to Eq. (1) accounting for Eq. (2) is presented in the mathematical appendix (see Supplementary Information). This solution accounts for the electromagnetic coupling with the magnetic shield explicitly in its construction. To relate this solution to the magnetic field, we define the azimuthal and axial currents on the coil in terms of a symmetric single-valued streamfunction decomposed in a Fourier basis, such that

$$J_\phi(\phi',z') = \frac{\partial \varphi(\phi',z')}{\partial z'}\,, \qquad J_z(\phi',z') = -\frac{1}{\rho_c}\frac{\partial \varphi(\phi',z')}{\partial \phi'}\,, \tag{3}$$

and

$$\varphi(\phi',z') = \frac{L_c}{n\pi}\left[\sum_{n=1}^{N} W_{n0}\cos\left(n\pi\left(\frac{z'}{L_c}-\frac{1}{2}\right)\right) - \sum_{n=1}^{N}\sum_{m=1}^{M}\left(W_{nm}\cos(m\phi') + Q_{nm}\sin(m\phi')\right)\sin\left(n\pi\left(\frac{z'}{L_c}-\frac{1}{2}\right)\right)\right]. \tag{4}$$

The Fourier coefficients, ($W_{n0}$, $W_{nm}$, $Q_{nm}$), are summed to $N$ and $M$, with higher $n$ and $m$ representing modes with greater axial and azimuthal spatial frequency, respectively. We can design a coil to generate any physical target field by matching equations (S.7–S.9) in the mathematical appendix at a dense set of points inside the target field region to the target profile. We employ a least-squares fitting to the desired magnetic field at the set of target points to find the optimal Fourier coefficients which generate the target field. By substituting the optimal Fourier coefficients into Eq. (4) and contouring over the full domain of the streamfunction with an appropriate number of contour levels, $N_{contours}$, we can generate discrete streamlines which represent the positions of wires on the coil. Increasing the number of levels reduces the difference between the field generated by the wires and by the continuum of current flow on the cylinder but may make the coil harder to manufacture.

**Design.** We now use the theoretical model to design a bespoke coil system for an existing quantum sensor—a gravimeter based on rubidium cold atom interferometry. In each measurement, a Mach–Zehnder atom interferometry sequence[19] is applied to a falling atomic cloud. This is illustrated in Fig. 2a, where the interferometer phase is encoded in the atomic ground state populations. In the ideal case, the phase shift is induced only by gravitational acceleration. However, any external force can induce a phase shift by displacing the atoms relative to the laser wavefronts. To operate the interferometer, a quantisation axis is defined with a bias field, and, for atoms with magnetic sub-levels, lifts the energy degeneracy via Zeeman splitting. This enables the interferometer to operate on transitions between the magnetically insensitive $m_F = 0$ sub-levels, where $m_F$ is the magnetic quantum number of the atomic state.

Schematics of the magnetic package of the interferometer and its magnetic systems are shown in Fig. 2b. The interferometer is enclosed in a three-layer capped cylindrical mumetal passive magnetic shield of internal radius $\rho_s = 110$ mm and length $L_s = 769$ mm. Each layer is constructed from 1 mm thick mumetal plate, which is rolled, welded, and then heat treated. The lower end cap contains entry holes for structural and electronic access. The system is constructed, wherever possible, of materials with no undesired magnetic properties. For example, the vacuum system is made of titanium, optics mounts are made of aluminium, and screws are made of low magnetic steel. The atomic package contains an ion pump (Gamma Vacuum 3S, 2018) that must remain inside the magnetic shielding and coil system. The ion pump has previously been shielded with a cuboidal mumetal magnetic to reduce the field from its strong magnet. The residual field inside the shielding varies depending on several factors, including the external background, leakage from the ion pump, and the orientation of the passive shields due to remnant magnetisation from extensive use. While the passive shielding is sufficient in ideal environments, further shielding is desired in a noisy background and to address internal magnetic field sources or magnetisation of the shield itself. To alleviate this, residual magnetic fields can be nulled using sets of simple rectangular compensation coils (see Supplementary Information). Here, we wish to design an integrated solution with greater performance than the simple system, generating magnetic fields more accurately and over a larger region. We want to control the total magnetic field in an extended optimisation region of radius $\rho_o = 10$ mm and length $L_o = 334$ mm. The optimisation region includes the entire region through which the atom cloud passes during preparation and while falling during interferometry.

We now use the theoretical model to design a coil system which is retrofitted inside the interferometer. We design six independent coils to generate six orthogonal magnetic fields in the optimisation region: three uniform fields ($B_x, B_y, B_z$) and three constant linear gradient fields with respect to the $z$-axis of the interferometer ($dB_x/dz, dB_y/dz, dB_z/dz$). These magnetic fields are selected because the optimisation region is long and thin, prioritising control of the field along the axis of the interferometer. The coils are wound on a single 3D-printed former of radius $\rho_c = 105$ mm and $L_c = 702$ mm (Fig. 3a–c). This former is designed to fit within the inner shield of the interferometer, close to the inside of the inner shield. By placing the coil close to the inner shield, we maximise the usable volume inside the shield and minimise errors due to inaccurate representation of the continuum current flows. There is a radial gap between the coil and shield to enable mounting and to ensure that the coils do not electrically short on the inner shield. As demonstrated theoretically in the mathematical appendix, the optimal uniform $B_z$ coil design when $L_c = L_s$ is a perfect solenoid. Because of this, since the coil former is only slightly shorter than the magnetic shield ($L_c/L_s = 0.913$), the uniform $B_z$ coil design is a solenoid of $N = 400$ turns, extending along the full length of the coil former. Current flows to generate the $B_x, B_z, dB_x/dz$, and $dB_z/dz$ fields are presented in Fig. 4.

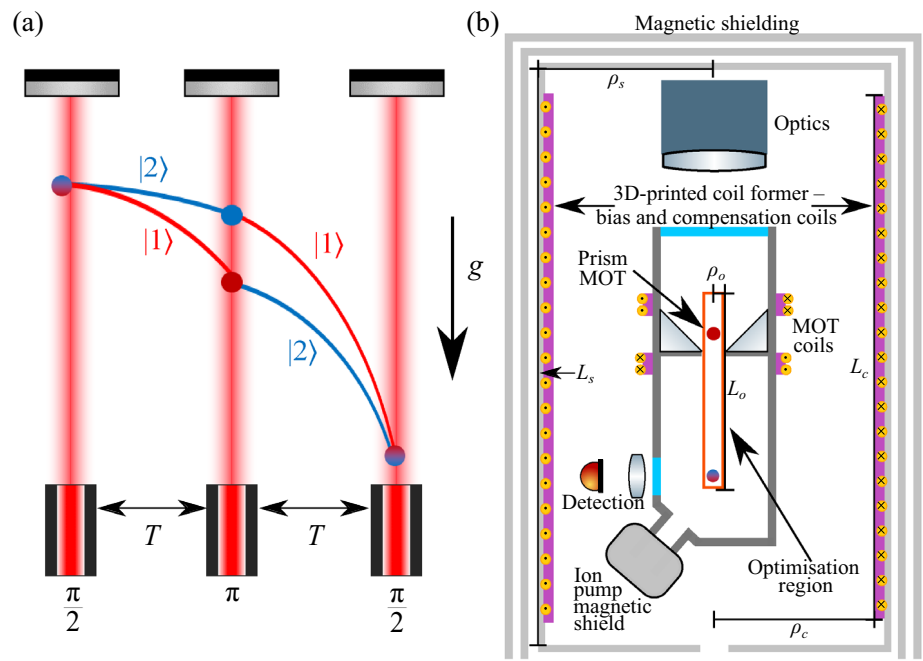

**Figure 2.** (**a**) Gravity, $g$, is measured using a Mach–Zehnder atom interferometry sequence[19]. The atom cloud is prepared in one of two atomic energy and momentum states, $|1\rangle$ and $|2\rangle$, and then is put into a quantum superposition of these states using stimulated Raman transitions. The states are allowed to evolve for a time $T$, before being swapped using a mirroring pulse and finally mixed after a further evolution time $T$. Due to the phase shift of the atom–light interaction, the sequence is sensitive to acceleration along the axis of the interrogation beams. These beams are aligned to the direction of gravitational acceleration. The value of gravity is proportional to the interferometer phase which is encoded in the state populations. (**b**) Schematic of the interferometer's magnetic shield and atomic measurement package, with the regions relevant to the specification of the coil system highlighted. The coil system sits inside the interferometer's magnetic shield and outside the measurement package. The coil system is designed to generate target fields in the optimisation region shown by the red rectangle.

## Characterisation.

The magnetic field coils are designed assuming that current flows as a continuum and the magnetic shielding is a closed perfect magnetic conductor. However, they are retrofitted as discrete coils into a magnetic shield with finite permeability and entry holes, which contains an atomic measurement package. To benchmark the theoretical model in this context, first, each coil is characterised with and without the magnetic shield, using a duplicate mounting structure of the interferometer which excludes the atomic package. These results are summarised in Table 1 and shown in Figs. 5 and 6.

In Fig. 5a the transverse field generated by the uniform $B_x$ coil is presented along the $z$-axis, with good agreement between the theoretical model and experimental data. The electromagnetic coupling with the shield enhances the spatial uniformity of the transverse field and amplifies its magnitude by a factor of 1.90 at the shield's centre. We can quantify the performance of the coil by calculating the difference between the measured and desired field profile along the $z$-axis of the optimisation region. The experimentally measured transverse field, inside the shield, has a Root-Mean-Square (RMS) deviation of 0.08% and a maximum deviation of 0.18%. This is over a hundredfold improvement in maximum field deviation when compared to the previously used rectangular compensation coil pair and over a threefold improvement when compared to a standard saddles adapted from Ref.[31] (see the Supplementary Information). Additionally, in Fig. 5b–d, we demonstrate that the transverse field is uniform along $z$-lines at the radial centre, half-way point, and edge of the optimisation region. In Fig. 5, we also use Finite Element Method (FEM) software to numerically simulate the discrete wires on the coil inside the real-world shield, excluding the atomic package, with good agreement along all measurement axes. Along the $z$-axis, the RMS deviation between the experimentally measured and the numerically simulated transverse field is 0.09%. The small deviations between the experiment and simulation likely result from differences in the shield permeability, small errors in the winding of the coil former, and wear-and-tear of the shield during previous use.

In Fig. 6a, the axial field generated by the uniform $B_z$ coil is presented along the $z$-axis. Again, we see that the axial field agrees closely with both the theoretical model and numerical simulation. In Fig. 6b, the uniformity of the axial field is presented along the $z$-axis. As expected from the model, the magnetic shield acts to increase the magnitude of the axial field by 1.04 at the shield's centre and improves the uniformity of the field; the RMS difference between the measured data and a uniform field is 0.98% without the magnetic shield and reduces to 0.33% when the magnetic shield is added.

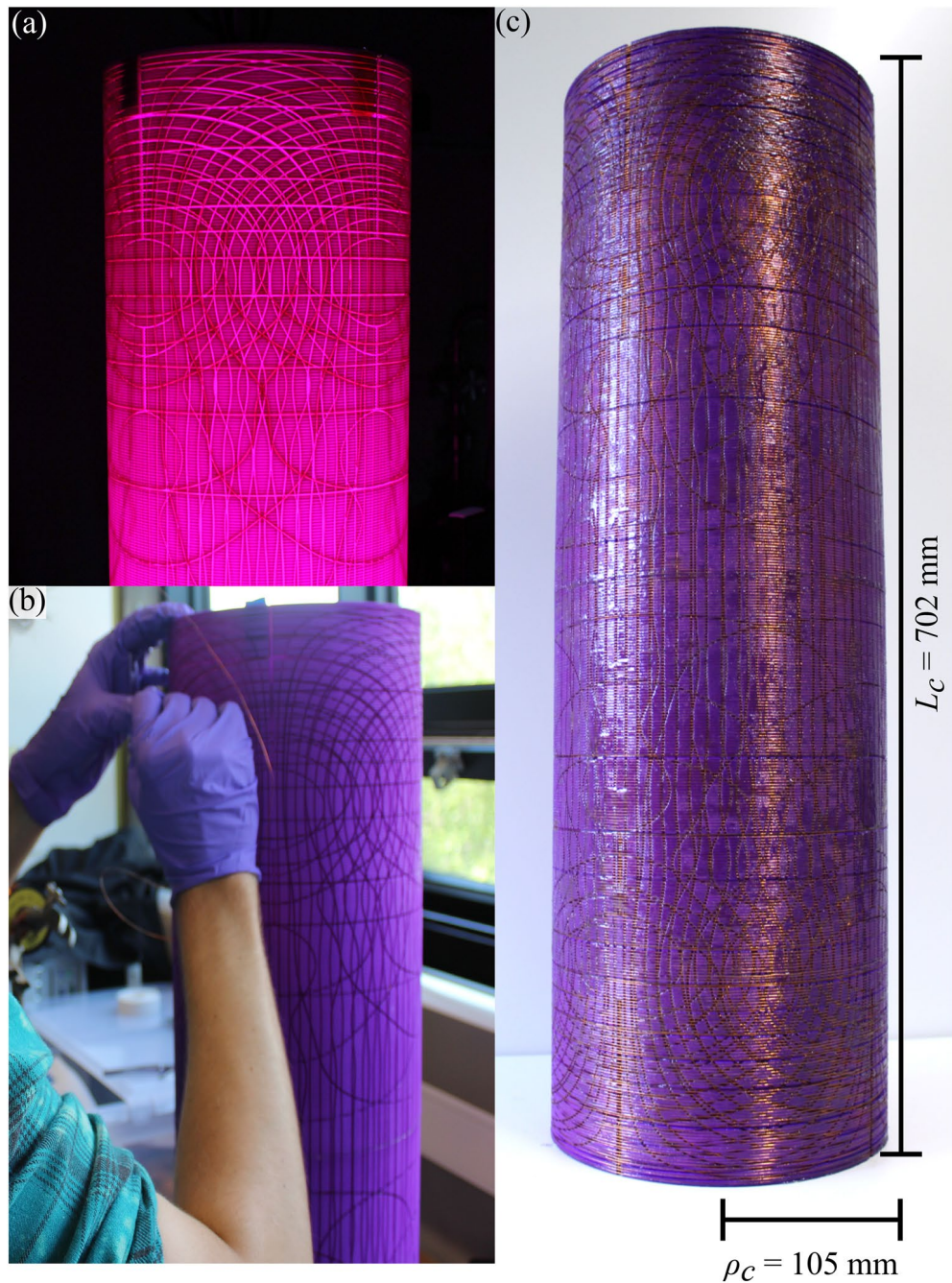

**Figure 3.** The 3D-printed coil former (**a**) before, (**b**) during, and (**c**) after winding.

**Atomic mapping.** We now measure the total magnetic field produced by the uniform $B_z$ coil with the Zeeman splitting of the magnetic sub-level resonances. This gives a direct measurement of the magnetic field profile experienced by the atoms during their drop. The uniform $B_z$ coil is selected for this measurement since it aligns with the axis of the interrogation beams. The coil is operated at a current of $I = 100$ mA, and the field, as measured by the atoms, is shown versus the vertical distance in the chamber in Fig. 7, alongside a numerical simulation of the field, including the effects of the ion pump magnetic shield. The field is measured as the atoms fall from the MOT at position $z = 68.1$ mm ($t = 0$) to the final interferometry pulse at position $z = -103.4$ mm ($t = 187$ ms). Analysing the experimental data in this region, the total bias field has an RMS deviation from perfect uniformity of 0.53%, whereas the bias field previously characterised without the ion pump has an RMS deviation of 0.08%. Clearly, the bias field uniformity is no longer limited by the coil topology or the distortion from the main shielding, and instead is limited by distortion from the ion pump shield. In the future, the numerical simulations may allow the coil designs to be perturbed to account for the effect of the ion pump shield in their topology.

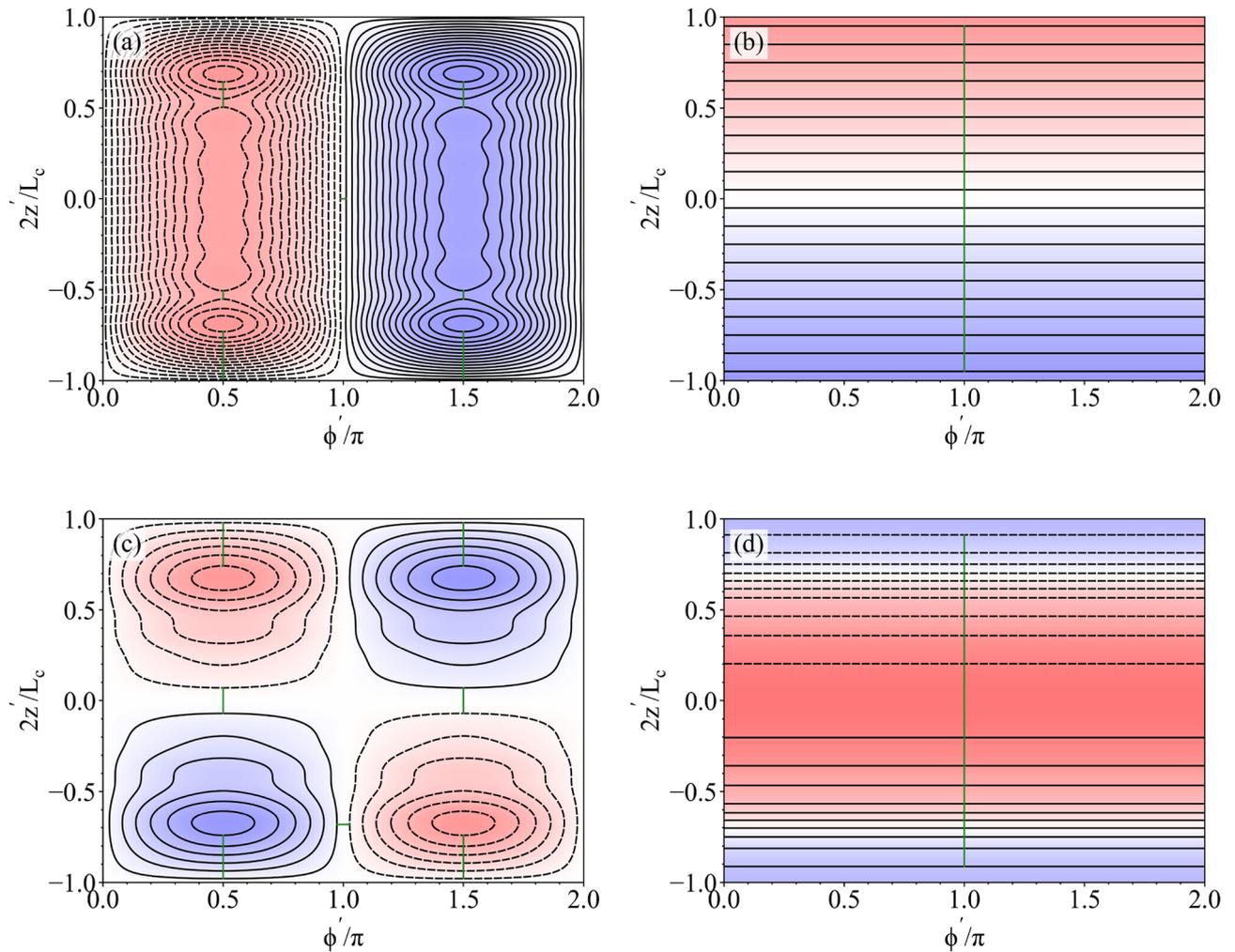

**Figure 4.** Normalised streamfunctions, calculated using Eq. (4), of the (**a**) uniform transverse field $B_x$, (**b**) uniform axial field $B_z$, (**c**) constant linear transverse field gradient $\mathrm{d}B_x/\mathrm{d}z$, and (**d**) constant linear axial field gradient $\mathrm{d}B_z/\mathrm{d}z$ coils, respectively, on the surface of the coil former of length $L_c$ (blue and red shaded regions correspond to the flow of current in opposite directions and the intensity shows the streamfunction magnitude). Each streamfunction is contoured into streamlines with solid and dashed black curves representing with opposite directions of current flow. To aid view, the uniform $B_z$ coil is represented with only 20 contour levels, whereas 400 are used in the real world design. Green solid wires show locations where the streamlines are joined with insulated entry and exit wires, following the direction indicated by the linestyles of the streamlines.

| $N$ | Field coil | Field profile | $N_{\text{contours}}$ | $l$ (m) | $B/I$ ($\mu$T/(Am$^{(N-1)}$)) | $\Delta B^{\text{RMS}}$ (%) | max($|\Delta B|$) (%) |
|---|---|---|---|---|---|---|---|
| 1 | $B_x$ | $\mathbf{B} = \hat{\mathbf{x}}$ | 30 | 44.8 | 160 | 0.08 | 0.18 |
| 1 | $B_z$ | $\mathbf{B} = \hat{\mathbf{z}}$ | 400 | 270.0 | 718 | 0.33 | 0.84 |
| 2 | $\mathrm{d}B_x/\mathrm{d}z$ | $\mathbf{B} = (z\,\hat{\mathbf{x}} + x\,\hat{\mathbf{z}})$ | 12 | 16.3 | 136 | 2 | – |
| 2 | $\mathrm{d}B_z/\mathrm{d}z$ | $\mathbf{B} = (-x\,\hat{\mathbf{x}} - y\,\hat{\mathbf{y}} + 2z\,\hat{\mathbf{z}})$ | 10 | 13.5 | 108 | 6 | – |

**Table 1.** Properties of the field coils when the coil system is shielded. Six independent coils generate three order $N = 1$ uniform fields ($B_x$, $B_y$, $B_z$) and three order $N = 2$ constant linear field gradients ($\mathrm{d}B_x/\mathrm{d}z$, $\mathrm{d}B_y/\mathrm{d}z$, $\mathrm{d}B_z/\mathrm{d}z$). The $B_x$ and $\mathrm{d}B_x/\mathrm{d}z$ coils have identical properties to the $B_y$ and $\mathrm{d}B_y/\mathrm{d}z$ coils, respectively. For each wire configuration of length $l$, the streamfunction is contoured with $N_{\text{contours}}$ to generate a magnetic field profile $B$ with a field strength per unit current $B/I$ averaged along the $z$-axis of the optimisation region. For each coil, the difference between the normalised field in the standalone shield and the target field, $\Delta B = B - B^{\text{des}}$, as a percentage of the desired field, is evaluated along the $z$-axis of the optimisation region and its root-mean-square (RMS) value is calculated. For the $B_x$ and $B_z$ coils, the maximum deviation from perfect field uniformity, max($|\Delta B|$), is also evaluated in the same region. All magnetic field values are calculated from experimentally measured data.

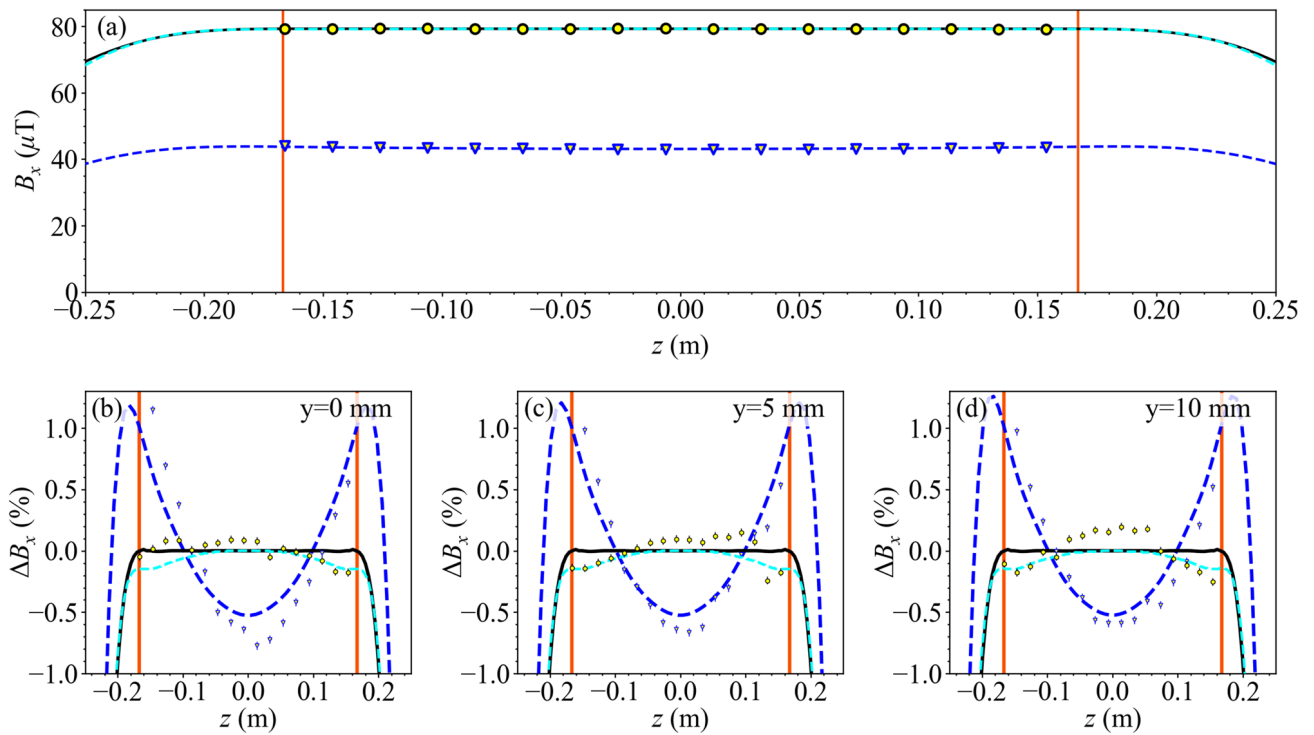

**Figure 5.** (**a**) Transverse magnetic field generated by the uniform $B_x$ coil (current $I = 500$ mA) with/without the magnetic shield (black/dark blue) evaluated along the $z$-axis. Scatter points (yellow circle/triangle with black/dark blue outline) show measured data and lines show theoretical field profiles calculated analytically from the theoretical model (solid black) or numerically via the Biot–Savart law (dashed dark blue). An additional numerical simulation shows the field generated by the discrete coil configuration inside the standalone shield, excluding the atomic package (dashed cyan; COMSOL Multiphysics Version 5.3a). Overlaid lines (solid red) show the edges of the optimisation region. (**b–d**) Deviation from perfect uniformity of the normalised transverse field, $\Delta B_x = B_x - 1$, along the $z$-axis, and displaced by $(x, y) = (0, 5)$ mm and $(x, y) = (0, 10)$ mm inside the optimisation region of radius $\rho_o = 10$ mm. Labelled identically to (**a**).

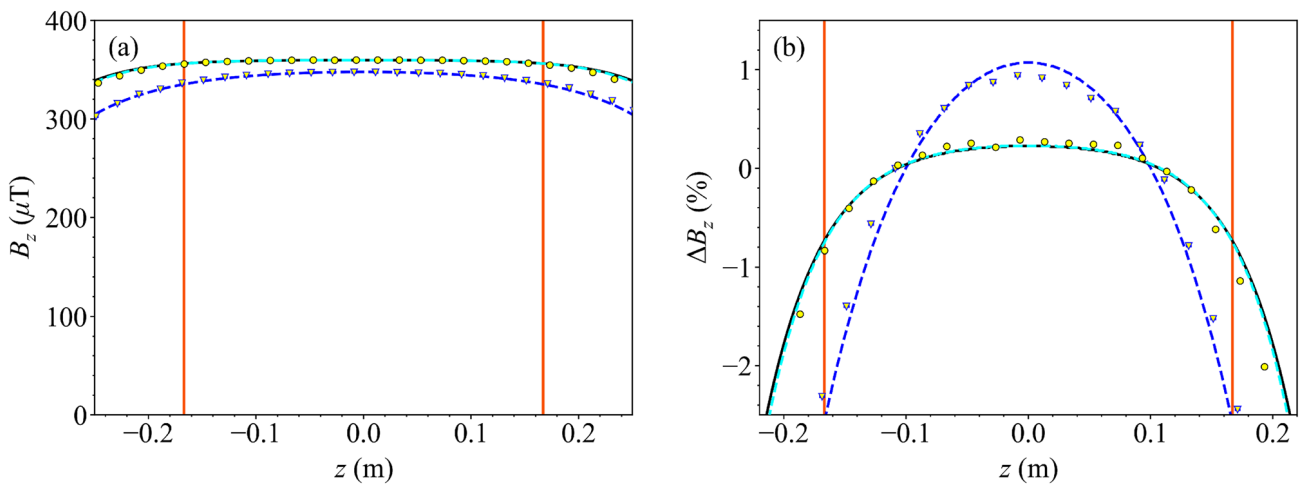

**Figure 6.** (**a**) Axial magnetic field generated by the uniform $B_z$ coil (current $I = 500$ mA) with/without the magnetic shield (black/dark blue) evaluated along the $z$-axis. Scatter points (yellow circle/triangle with black/dark blue outline) show measured data and lines show theoretical field profiles calculated analytically from the theoretical model (solid black) or numerically via the Biot–Savart law (dashed dark blue). An additional numerical simulation shows the field generated by the discrete coil configuration inside the standalone shield, excluding the atomic package (dashed cyan; COMSOL Multiphysics Version 5.3a). Overlaid lines (solid red) show the edges of the optimisation region. (**b**) Deviation from perfect uniformity of the normalised axial field, $\Delta B_z = B_z - 1$, along the $z$-axis. Labelled identically to (**a**).

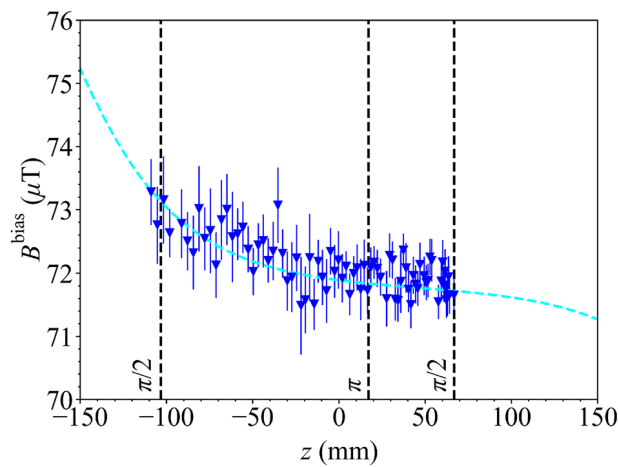

**Figure 7.** The total bias magnetic field, $B^{bias}$, generated by the uniform $B_z$ coil evaluated along the $z$-axis of the interferometer using the atoms (dark blue scatter points), and fitted to numerical simulation of the field profile in the interferometer with the atomic package (dashed cyan; COMSOL Multiphysics Version 5.3a). Additional lines (dashed black) show the axial position of the cloud during the interferometry pulses (dashed black; in $\pi/2 - \pi - \pi/2$ sequence at distances $z = [66.9, 17.1, -103.4]\,\text{mm}$) as the cloud falls from the MOT ($z = 68.1\,\text{mm}$).

**Phase error reduction.** As we have matched the numerical simulations to the atomic map, we can use the numerically simulated field profiles to calculate the theoretical benefit of the coil system to the interferometer via reducing the quadratic Zeeman effect induced phase shift. As an example, let us consider an ambient field containing the bias field numerically simulated above, plus a uniform background transverse field component of $B^{BG} = 100\,\text{nT}$, observed during operation outside of the laboratory to cause unwanted phase shifts in the instrument. When compensating the transverse field using the uniform $B_x$ coil rather than the standard rectangular coil (see Supplementary Information), there is a theoretical reduction in the systematic shift from the quadratic Zeeman effect of nearly four orders of magnitude, from 90 mrad to 9.9 µrad.

## Discussion

In this work, we have designed multiple current-carrying coils on a 3D-printed coil former in a cold atom interferometer which generate three uniform and three constant linear gradient magnetic fields, while accounting for the response of the interferometer's external high permeability magnetic shield. The theoretical model presented in the text accurately predicts the shape of the field, as shown by good agreement between the model, simulation, and the in-situ experimental data. A uniform transverse field is generated which deviates by less than 0.2% over more than 40% of the length of the sensor shield. We also theoretically and experimentally demonstrated that the field generated by a solenoid is improved via electromagnetic coupling with the shield. In the future, to reduce magnetic field deviations even further, nested flexible PCBs[45] may enable coil streamlines to be represented more densely and accurately. This may enable the construction of bias coil designs with higher frequency spatial variations in the current flow patterns or further magnetic field coils to be added.

We then measured how the internal atomic package affected the magnetic field profile and theoretically calculated the reduction of quadratic Zeeman effect induced phase shift via optimal magnetic field control. Future research may consider the effect of using magnetic field control with wave vector reversal to further mitigate the quadratic Zeeman effect, by simulating the atom fall trajectory between reverses of the wave vector in a noisy magnetic field environment and when this environment is controlled using a coil system. Moving forward, active cancellation of multiple magnetic field components using the coil system[46,47] will be implemented using a redesigned atomic package with spatially separated magnetometers positioned directly around the optimisation region.

Coils can be designed quickly using our compact formulation and may be retrofitted to existing systems. This makes our method applicable to a diverse range of settings, from small compact field sensors[4] to extremely large drop towers[48] aimed at testing fundamental physics. When used in combination with in-built degaussing systems and optimal passive shield arrangements[49], our method offers improved shielding efficiency and a significant reduction in weight and cost.

## Methods

### Design and manufacture of the coil former.
To design the former, discrete representations of the streamfunctions in Fig. 4 are simulated numerically using FEM software, following the method in Ref.[50]. The streamlines are selected to minimise both the maximum deviation between the target field and desired field along the $z$-axis of the optimisation region and the number of overlaps between different wire configurations. The representation of the uniform field coils is prioritised. Each set of streamlines to generate each target field is then joined, at the points where the streamlines are closest, to make a continuous wire path (Fig. 4). When joining the streamlines, the return paths are traced back over with the opposite current polarity to cancel the erroneous magnetic fields generated by the connections.

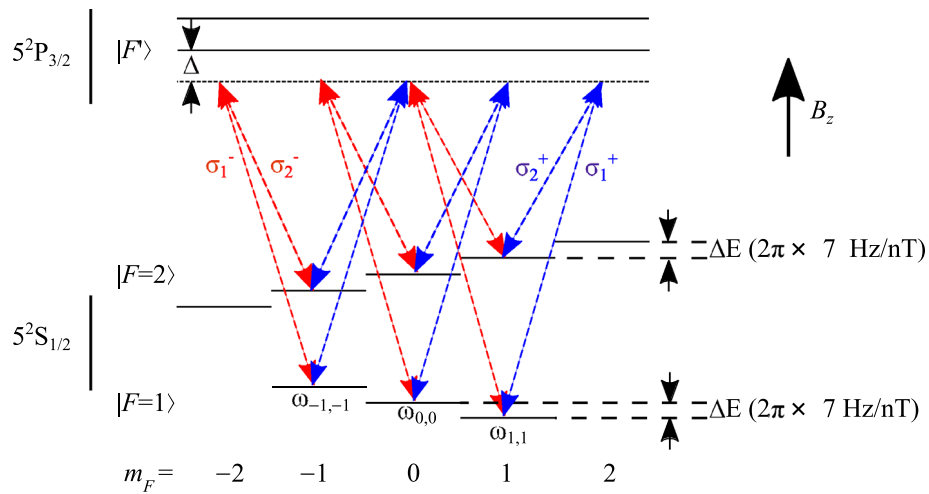

**Figure 8.** The $^{87}$Rb ground state magnetic sub-levels and possible Raman transitions using the $D_2$ line of angular frequency $\omega$. $|F\rangle$ and $|F'\rangle$ represent the ground and excited state energy levels, respectively, and $m_F = 0, \pm 1, \ldots, \pm F$ are the projections of the total angular momentum $F$ on to the quantisation axis. The red and blue arrows show the possible transitions for Raman beams with $\sigma^+/\sigma^-$ polarised light with a uniform axial magnetic field, $B_z$, applied parallel to the Raman beam. Only three Raman transitions are allowed, with frequencies $\omega_{-1,-1}$, $\omega_{0,0}$, and $\omega_{1,1}$.

The 3D model of the former was designed using SOLIDWORKS. The grooves for coils of different orders (see $N$ in Table 1) are designed at different radial depths on the former. The 3D-printed former is presented in Fig. 3a. It is printed by Selective Laser Sintering (SLS) 0.12 mm thick layers of nylon using an EOS P760 SLS machine (3T-AM LTD, 2019). The former is hand-wound with 0.45 mm diameter enamelled copper wire, allowing a maximum current of 1.0 A. The output wires are wrapped together as twisted pairs. The wound coil system weighs 2.58 kg, whereas the existing magnetic shields for the system weigh 23.83 kg.

**Characterising the coil system.** The magnetic fields are characterised using a standalone mounting structure of the interferometer. Each measurement is taken by averaging the reading of a magnetometer for 30 s (Fluxgate Fluxmaster, Stefan Mayer Instruments 2018) both with and without the interferometer magnetic shield. To maximise the signal to noise, characterisation is performed with a large field magnitude relative to the field background, which is subtracted from each measurement. The standard errors in the measured magnetic field at each point are negligible. Greater errors are introduced due to the misalignment of the coil, shield, and magnetometer, particularly when the shields are removed from the system to move the magnetometer. Repeated measurements of the transverse magnetic field generated by the uniform $B_x$ coil (Fig. 5), in the optimisation region, where re-alignment is performed between each repeat, show deviations at the 0.01% level. This is assumed to be the error in the calculated RMS and maximum errors in the fields generated by the uniform field coils. The characterisation of the gradient coils is detailed in the Supplementary Information.

**Mapping the magnetic field using cold atoms.** The magnetic field within the interferometry chamber is mapped using Raman spectroscopy-based magnetic field mapping using the $D_2$ transition of $^{87}$Rb, as described in Ref.[17]. The three allowed Raman transitions for the $D_2$ transition of $^{87}$Rb under the application of an axial bias field are presented in Fig. 8. The transition frequencies, $\omega_{1,1}$, $\omega_{0,0}$, $\omega_{1,-1}$, are found by Lorentzian fitting[51] to the Raman spectrum as presented in Fig. 9a.

Having measured these resonance transition frequencies, the magnetic field at the position of atom–light interaction can be inferred by considering the frequency difference between the transitions,

$$B_1 = (\omega_{1,1} - \omega_{0,0})/\gamma_1 \,, \tag{5}$$

$$B_{-1} = (\omega_{0,0} - \omega_{-1,-1})/\gamma_1 \,, \tag{6}$$

where $\gamma_1 = 2\pi \times 14$ Hz/nT is the first-order Zeeman coefficient[52]. The mean of $B_1$ and $B_{-1}$ is the total magnetic flux density, which consists of the total magnetic field generated by the uniform $B_z$ coil, $B^{\text{bias}}$, and the total background magnetic field, $B^{\text{BG}}$. To separate these fields, the magnetic field is mapped within the chamber at the full bias coil current ($I = 100$ mA), where $B = B^{\text{bias}} + B^{\text{BG}}$, and the half bias coil current ($I = 50$ mA), where $B = 0.5B^{\text{bias}} + B^{\text{BG}}$. Field maps at these currents are presented in Fig. 9b–c. This method is applied to generate a magnetic field map along the $z$-axis at $t = 1$ ms intervals during free-fall of the atomic cloud. The Raman spectra are measured at each height by applying Raman pulses to the clouds with a different detuning each time.

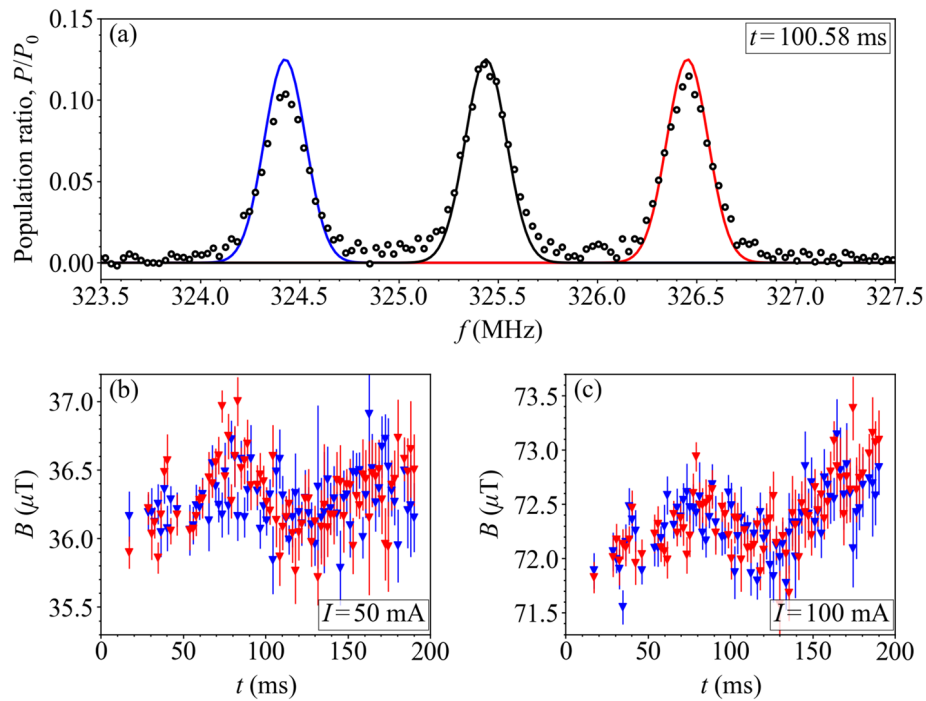

**Figure 9.** (**a**) Population ratio, $P/P_0$, of the $^{87}$Rb $5^2S_{1/2}$ $|F = 1\rangle$ ground state as a function of Raman frequency, $f$, at a drop time $t = 100.58$ ms (position $z = 18.5$ mm), with a Raman pulse length of 4 μs and frequency step of 30.30 kHz under axial magnetic field generated by the uniform axial $B_z$ field coil at current $I = 50$ mA. The data is shown with 133 points and the Lorentzian fits to the $\omega_{-1,-1}$, $\omega_{0,0}$, $\omega_{1,1}$ resonances are highlighted (blue, black, red). (**b,c**) Measured total magnetic field, $B$, calculated from the $B_{-1}$ (blue) and $B_1$ (red) transitions scanned at 78 points during the atomic cloud drop at bias coil currents: (**b**) $I = 50$ mA and (**c**) $I = 100$ mA.

**Quadratic Zeeman effect.** Following Ref.[17], we calculate the phase shift induced from the quadratic Zeeman effect as

$$\theta^{\text{Zeeman}} = \gamma_2 \int_0^{2T} dt \, g_s(t) B^2(t) , \qquad (7)$$

where $\gamma_2 = 2\pi \times 0.0575$ Hz/nT$^2$ is the second-order Zeeman coefficient[52], $g_s(t)$ is the sensitivity function of the interferometer[53], $T$ is the pulse separation, and $B(t)$ is the total magnetic field at a time $t$ during the atomic cloud fall from $t = 0$ to $t = 2T$. During typical operation of the interferometer, $T = 85$ ms and the pulse widths of the $\pi/2$ and $\pi$ pulses are very short, $\tau_{\pi/2} = 2$ μs and $\tau_\pi = 4$ μs, respectively, meaning that the phase shift only needs to be evaluated in-between the pulses.

## Data availability

All supporting data is available upon request.

## Acknowledgements

We acknowledge support from EPSRC through grants EP/M013294/1 and EP/T001046/1, Innovate UK through projects 44430 and 104613, and the Ministry of Defence, as part of the UK National Quantum Technologies Programme. The authors would also like to acknowledge the physics workshop team at the University of Birmingham for machining several parts essential to the experimental systems used in this work.

## Author contributions

J.V., P.J.H., M.H., B.S. and K.B. and conceived and designed the experiments. M.P., P.J.H., N.H., M.J.B., R.B., and T.M.F. developed the magnetic field optimisation procedure. P.J.H., M.P., J.V., and B.S. designed the coil former. P.J.H. and J.W. constructed the coil. P.J.H., J.V., J.W., K.M. and B.S. acquired the data. B.S., J.V., J.W., F.H., and M.H. constructed and operated the interferometer used in these experiments. P.J.H., J.V., B.S., and J.W. performed the data analysis. All authors drafted and revised the manuscript.

## Competing interests

The authors M.P, P.J.H, T.M.F, M.J.B, and R.B declare that they have a published worldwide patent (WIPO Patent Application WO/2021/053356) including the magnetic field optimisation techniques described in this work. N.H, M.J.B, and R.B also declare financial interests in a University of Nottingham spin-out company, Cerca. The authors have no other competing financial interests. The authors declare that they have no non-financial competing interests.

## Additional information

**Supplementary Information** The online version contains supplementary material available at https://doi.org/10.1038/s41598-022-13979-4.

**Correspondence** and requests for materials should be addressed to T.M.F.

**Reprints and permissions information** is available at www.nature.com/reprints.

**Publisher's note** Springer Nature remains neutral with regard to jurisdictional claims in published maps and institutional affiliations.